\documentclass{elsart}
\usepackage{amssymb,amsmath,cite,epsfig} 

\def\b{\beta}                   
\def\g{\gamma}                  
\def\e{\epsilon}                

\def\wt{\widetilde}

\newcommand{\bea}{\begin{eqnarray}}
\newcommand{\eea}{\end{eqnarray}}

\font\tengoth=eufm10 \font\sevengoth=eufm7 \font\fivegoth=eufm5
\newfam\gothfam
\textfont\gothfam=\tengoth \scriptfont\gothfam=\sevengoth
  \scriptscriptfont\gothfam=\fivegoth

\begin{document}

\begin{frontmatter}

\title{Second order SUSY transformations \\ with `complex energies'}

\author[mex]{David J. Fern\'andez C.\thanksref{corrauth}},
\ead{david@fis.cinvestav.mx}
\author[mex]{Rodrigo Mu\~noz}
\ead{rodrigom@fis.cinvestav.mx}
and
\author[mex,pad]{Arturo Ramos\thanksref{citaart}}
\ead{aramos@math.unipd.it}
\address[mex]{Departamento de F\'{\i}sica, CINVESTAV, A.P. 14-740, 07000
M\'exico D.F., Mexico}
\address[pad]{Dipartimento di Matematica Pura ed Applicata, Universit\`a
degli Studi di Padova, Via G. Belzoni 7, I-35131 Padova, Italy}
\thanks[corrauth]{Corresponding author. Tf: +52 57473879. Fax:
+52 57473879.}

\thanks[citaart]{Work developed in part while A.R. was visiting the Dept. 
de F\'{\i}sica of CINVESTAV in the fall of 2000. On leave of absence from
Departamento de F\'{\i}sica Te\'orica, Facultad de Ciencias,
Universidad de Zaragoza, 50009 Zaragoza, Spain.}

\begin{abstract}
\noindent Second order supersymmetry transformations which involve a pair
of complex conjugate factorization energies and lead to real non-singular
potentials are analyzed. The generation of complex potentials with real
spectra is also studied. The theory is applied to the free particle,
one-soliton well and one-dimensional harmonic oscillator. 
\end{abstract}

\begin{keyword} 
Second-order supersymmetry\sep irreducible intertwining operators\sep
complex potentials with real spectra\sep generation of solvable potentials

\PACS 03.65.Ca\sep03.65.Fd\sep03.65.Ge\sep11.30.Pb
\end{keyword}
\date{23 August 2002}
\end{frontmatter}

\section{Introduction}

The $n$-th order supersymmetric quantum mechanics ($n$-SUSY QM), which
involves differential intertwining operators of order $n$, is a useful
tool for generating new solvable potentials
\cite{AndIofSpi93,BagSam97,FerHusMiel98,FerHus99}. Due to its simplicity,
the 1-SUSY QM is the most explored; its nonsingular transformations
produce partner potentials whose spectra can differ at most in the ground
state energy level \cite{CarMarPerRan98,CarRam00b,CarRam00c}. The
difficulty of `modifying' the excited part of the spectrum has been
surpassed through the 2-SUSY QM
\cite{AndIofCanDed95,AndIofNis95,Fer97,FerGlaNi98,Ros98_2,MieNieRos00,
CarRamFer01}, which allows to `create' two new levels $\epsilon_1, \
\epsilon_2$ between two neighboring energies $E_i, \ E_{i+1}$ of the
initial Hamiltonian \cite{Sam99}.  A similar treatment, implemented for
periodic potentials \cite{FerNegNie00}, can be used to embed two bound
states in a spectral gap above the lowest energy band
\cite{FerMieRosSam02,FerMieRosSam02b}. In both situations the
corresponding 2-SUSY transformations are {\it irreducible}, i.e., when
obtained as the iteration of two 1-SUSY procedures they will involve
always singular intermediate potentials. 

Here we will study a different set of `irreducible' 2-SUSY transformations
applied to non-periodic potentials, which employs two complex conjugate
factorization energies $\epsilon_1, \ \epsilon_2$,
$\epsilon_2=\bar\epsilon_1$.  Irreducibility means now that the
intermediate potential is complex although the final one is real. This
problem has been addressed previously \cite{BaOvSa95}, but up to our
knowledge the conditions granting that the final potential will be regular
have not been yet examined. We will show as well that the non-singular
case leads to intermediate complex potentials having real spectra. These
points constitute the subject of this letter, which has been organized as
follows.  In section 2 the second order SUSY transformations with
$\epsilon_1,\epsilon_2\in{\mathbb C}$, $\epsilon_2=\bar\epsilon_1$ will be
analyzed. A prescription for avoiding the singularities in the new
potential will be provided in section 3, while section 4 will be devoted
to study the intermediate complex potentials. Section 5 will deal with
some particular examples as the free particle, one-soliton well and
harmonic oscillator potential. 

\section{Second order supersymmetric quantum mechanics\label{fst_ord_int}}

The standard supersymmetric (SUSY) algebra with generators $Q_1, \ Q_2$
(supercharges) and $H_{\rm ss}$ (SUSY Hamiltonian) reads: 
\bea
& \{ Q_j,Q_k\} = \delta_{jk} H_{\rm ss}, \quad [H_{\rm ss},Q_j]=0, \quad
j,k=1,2, \label{susyal}
\eea
where $[\cdot,\cdot]$ denotes a commutator and $\{\cdot,\cdot\}$ an
anticommutator. The realization $Q_1 =(Q^\dagger + Q)/\sqrt{2}, \ Q_2 =
(Q^\dagger - Q)/(i\sqrt{2})$ with
\bea
& Q = \left(\begin{matrix} 0 & A \\ 0 & 0 \end{matrix}
\right), \quad Q^\dagger = \left(\begin{matrix} 0 & 0 \\ A^\dagger & 0
\end{matrix}
\right), \\
& H_{\rm ss} =
\left(\begin{matrix} A A^\dagger & 0 \cr 0 & A^\dagger A 
\end{matrix} \right) = \left(\begin{matrix}
(\wt H - \epsilon_1)(\wt H - \epsilon_2) & 0 \\ 0 & (H - \epsilon_1)(H -
\epsilon_2) \end{matrix}\right), \label{factorized} \\
& A = \frac{d^2}{dx^2}+\eta(x)\frac{d}{dx}+\g(x)\ ,
\label{defA_seg_ord}
\eea
is called second order sypersymmetric quantum mechanics ($2$-SUSY QM). In
this formalism $H, \ \wt H$ are two intertwined Schr\"odinger
Hamiltonians: 
\bea
& \wt HA = AH,  \label{intertwining} \\
& H=-\frac{d^2}{dx^2}+V(x), \quad
\wt H=-\frac{d^2}{dx^2}+\wt V(x)\, ,
\label{defHHtil} 
\eea
and thus the real functions $\eta(x), \ \g(x)$ are related with $V(x), \
\wt V(x)$ through: 
\bea
\wt V-V&=&2 \eta^\prime,	\label{Vt_V_eta_prime_a1}\\
(\wt V-V)\eta&=&2 V^\prime+2 \g^\prime+\eta^{\prime\prime},	    
				\label{Vt_V_gamma_prime_eta_pprime_a1}\\
(\wt V-V)\g&=&\eta V^\prime+V^{\prime\prime}+\g^{\prime\prime}\,. 
				\label{Vt_V_V_prime_V_pprime_a1}
\eea
To decouple this system, substitute (\ref{Vt_V_eta_prime_a1}) in
(\ref{Vt_V_gamma_prime_eta_pprime_a1}) and integrate to obtain
\bea
& \g=d - V+\eta^2/2-\eta^\prime/2\,,
\label{gam_eta}
\eea
where $d\in{\mathbb R}$ is a constant. By plugging
(\ref{Vt_V_eta_prime_a1},\ref{gam_eta})  into
(\ref{Vt_V_V_prime_V_pprime_a1}), multiplying the result by $\eta$ and
performing the next integration one arrives to:
\bea
& \eta \eta^{\prime\prime}-(\eta^{\prime})^2/2
+2\eta^2\bigg(\eta^2/4 - \eta^{\prime} - V + d \bigg) + 2 c=0\,,
\label{eq_eta} 
\eea
$c\in{\mathbb R}$ being another constant. This formalism is useful for
generating new solvable potentials. To illustrate it, suppose that $c,d$
are fixed and $V(x)$ is an initial exactly solvable potential. The
complete determination of $\wt V(x)$ in (\ref{Vt_V_eta_prime_a1}) requires
to find solutions $\eta(x)$ of the nonlinear second order differential
equation (\ref{eq_eta}). Let us find them using the {\it Ans\"atz}
\cite{Ros98_2}:
\bea
& \eta^{\prime}(x)=\eta^2(x)+2\beta(x)\eta(x)-2\xi(x)\,, 
\label{ricc_eta}
\eea
where $\beta(x)$ and $\xi(x)$ are to be determined. After substituting
(\ref{ricc_eta}) into (\ref{eq_eta}) we obtain a system of equations from
which it follows that $\xi^2=c$. The essential part of the system is the
Riccati equation:
\bea
& \b^\prime+\b^2=V-\epsilon, \quad \epsilon = d + \xi.
\label{eqRicbeta}
\eea
As there exist two possible values of $\epsilon$, $\epsilon_1 = d +
\sqrt{c}$ and $\epsilon_2 = d - \sqrt{c}$ (which coincide with the
factorization energies in (\ref{factorized})), we indeed are dealing with
two equations (\ref{eqRicbeta}) whose solutions will be denoted by
$\beta_1(x)$, $\beta_2(x)$. This leads to a natural classification of the
2-SUSY transformations based on the sign of $c$, which will be discussed
elsewhere. Here, we restrict ourselves to the {\it complex} case for which
$c<0$ and then $\epsilon_1, \epsilon_2 \in {\mathbb C}$, $\epsilon_2 =
\bar\epsilon_1$. Since $V$ is real we can take $\b_2(x) = \bar\b_1(x)$,
i.e., the problem reduces to solve the Riccati equation for $\b_1$. In
addition, using (\ref{ricc_eta})  we get two equivalent expressions for
the real $\eta(x)$: 
\bea
\eta^{\prime}&=&\eta^2+2 \b_1\eta - 2i{\rm Im}(\e_1)\,,
\label{eta_beta_p}\\
\eta^{\prime}&=&\eta^2+2 \bar\b_1\eta+2i{\rm Im}(\e_1)\,. 
\label{eta_beta_m}
\eea
By subtracting both equations and solving for $\eta$, we obtain:
\bea
& \eta={\rm Im}(\e_1)/{\rm Im}(\b_1)\,.
\label{eta_compl}
\eea
Once we know $\eta$, the 2-SUSY partner potential $\wt V(x)$ is calculated
using
\bea
& \wt V =
V + 2\left[{\rm Im}(\e_1)/{\rm Im}(\b_1)\right]'\,.
\label{tildevcc}
\eea
Let us remark that the case we are dealing with has been previously
explored \cite{AndIofCanDed95}. However, we have not found any previous
analysis on how to avoid the singularities in $\wt V(x)$, a phenomenon
which seems almost unavoidable in the complex case \cite{BaOvSa95}. 

\section{The non-singular 2-SUSY potentials}

The simplest algorithms departing from and arriving at the exactly
solvable potentials should avoid the singularities which might appear in
$\wt V(x)$. Notice that a singular $\wt V(x)$ could be treated as a
non-singular partner potential of $V(x)$ in a restricted $x$-domain. 
However, this would require at the end to solve the initial Schr\"odinger
equation with modified boundary conditions loosing, in general, the
solvability of $H$ \cite{MarNegNie98}.

Let us rewrite first the formulae of section 2 in terms of solutions
$u_1(x)$ of the Schr\"odinger equation arising from (\ref{eqRicbeta}) by
the change $\beta(x) = u'(x)/u(x)$ \cite{CarRamFer01}: 
\bea
& - u''(x) + V(x) u(x) = \epsilon u(x). \label{schro}
\eea
Hence $\eta(x) = - 2i{\rm Im}(\epsilon_1)  \vert u_1\vert^2/W(u_1,\bar
u_1)$, where $W(u_1,u_2)  =u_1 u_2' - u_2 u_1'$ denotes the Wronskian of
$u_1$, $u_2$. From now on it is convenient to work with the real {\it
normalized Wronskian} $w(x) \equiv W(u_1,\bar u_1)/[2i{\rm
Im}(\epsilon_1)]$.  Therefore: 
\bea
& w'(x) = \vert u_1(x)\vert^2, \label{derpos} \\
& \eta(x) = - w'(x)/w(x), \\
& \wt V(x) = V(x) - 2 [w'(x)/w(x)]'. \label{tildevccsch}
\eea
In order that $\wt V(x)$ be non-singular, $w(x)$ must be nodeless. Since
$w(x)$ is an increasing monotonic function (see (\ref{derpos})), the
arising of zeros is avoided if
\bea
& \lim_{x\rightarrow\infty} w(x) = 0 \quad
{\rm or} \quad
\lim_{x\rightarrow-\infty} w(x) = 0.
\label{goodwm}
\eea
To ensure this requirement it is sufficient that either
\bea
& \lim_{x\rightarrow\infty} u_1(x) = 0 \quad 
{\rm or} \quad
\lim_{x\rightarrow-\infty} u_1(x) = 0.
\label{goodum} 
\eea
Such solutions are appropriate for generating non-singular potentials $\wt
V(x)$. Notice that a similar treatment can be designed for systems defined
in a generic interval $x \in (a,b) \subset {\mathbb R}$ by identifying in
(\ref{goodwm}-\ref{goodum}) $-\infty$ with $a$ and $\infty$ with $b$. 

\section{Complex potentials with real spectrum}

Although in principle irreducible, let us decompose the non-singular
2-SUSY transformations of the previous section into two 1-SUSY steps:
\bea
& \wt H A_2 = A_2 H_1, \quad
H_1 A_1 = A_1 H, 
\label{interaux1}
\eea
where
\bea
& H_1 = - \frac{d^2}{dx^2} + V_1(x) , \quad A_i = \frac{d}{dx} +
\alpha_i(x), \ i=1,2.
\label{inth1}
\eea
The 1-SUSY treatment implies that $\alpha_1, \ \alpha_2$ obey the Riccati
equations: 
\bea
& -\alpha_1' + \alpha_1^2 = V(x) - \epsilon_1, \label{riccatiaux1} \\
& -\alpha_2' + \alpha_2^2 = V_1(x) - \bar\epsilon_1, \label{riccatiaux2}
\eea
where $V_1(x) = V(x) + 2\alpha_1'$ and thus $\wt V(x) = V(x) + 2(\alpha_1
+ \alpha_2)'$. A simple comparison of (\ref{eqRicbeta}) with
(\ref{riccatiaux1}) leads to
\bea
& \alpha_1(x) = - \beta_1(x) = - u_1'(x)/u_1(x),
\eea
$u_1(x)$ being a solution of (\ref{schro}) behaving asymptotically as in
(\ref{goodum}). Moreover, by expanding $A = A_2 A_1$ and comparing the
result with (\ref{defA_seg_ord}) we find that:
\bea
& \alpha_2 = - \alpha_1 + \eta = \beta_1 + (\epsilon_1 -
\bar\epsilon_1)/(\beta_1 - \bar\beta_1). \label{fda}
\eea
This is a particular case of the finite difference B\"acklund algorithm
\cite{FerHusMiel98,MieNieRos00,CarRamFer01}, which algebraically
determines a solution to (\ref{riccatiaux2}) in terms of solutions of
(\ref{riccatiaux1}) for two different factorization energies (here
$\epsilon_1$ and $\bar\epsilon_1$). It is interesting as well to factorize
the involved Hamiltonians: 
\bea
& H = A_1^- A_1 + \epsilon_1, \quad H_1 = A_1 A_1^- + \epsilon_1,
\label{complexfactorized1} \\
& H_1 = A_2^- A_2 + \bar\epsilon_1, \quad \wt H = A_2 A_2^- +
\bar\epsilon_1,
\label{complexfactorized2} 
\eea
$A_i^- = -d/dx + \alpha_i(x), \ i=1,2$. Since $\alpha_1, \alpha_2$ and
$\epsilon_1$ are complex, the $A_i^-$'s are not adjoint to the $A_i$'s but
$A_i^\dagger = - d/dx + \bar\alpha_i(x), \ i=1,2$.

It is clear now that the complex intermediate potential $V_1(x)$ is
non-singular:
\bea
& V_1(x) = V(x) - 2[u_1'(x)/u_1(x)]' .
\label{complexpot}
\eea
To analyse the normalizability of the corresponding eigenfunction
associated to $E_n$,
\bea
& \psi^1_n(x) = c_n A_1\psi_n= c_n [\psi_n'(x) - u_1'(x)\psi_n(x)/u_1(x)],
\label{eigencomplex}
\eea
we will employ the operator relationship:
\bea
& \eta A_1 = H - \epsilon_1 + A.
\eea
{}From the validity of (\ref{goodum}) and the assumption of 
$\vert\vert\psi_n\vert\vert=1$, it turns out that $\wt\psi_n = A\psi_n /
\vert E_n - \epsilon_1\vert$ is normalized, and therefore the function
$\eta A_1\psi_n = (E_n - \epsilon_1)\psi_n + \vert E_n - \epsilon_1\vert
\wt\psi_n$ is normalizable as well. Thus, for $A_1\psi_n$ to be 
normalizable it is neccessary that $\eta^{-1}$ does not destroy the
normalizability of $(E_n - \epsilon_1)\psi_n + \vert E_n - \epsilon_1\vert
\wt\psi_n$. If this is the case (and this will happen for the examples we
discuss below), we obtain a complex potential $V_1(x)$ with real
eigenvalues $E_n$ \cite{CanJunTro98,AndIofCanDed99,BagMalQue01}. Let us
remark that complex Hamiltonians with real spectra have been studied
recently in the context of PT-symmetry and pseudo-Hermiticity \cite{BenBoe98,Mos02}. 

\section{Illustrative examples} 

We shall show that the previous techniques admit very simple applications. 

\smallskip

{\it i)} Consider firstly the free particle for which $V(x)=0$. The
general solution $u_1(x)$ of the Schr\"odinger equation (\ref{schro}) for
$\epsilon_1\in{\mathbb C}$ is a linear combination of
\bea 
& e^{\pm(k_1 + i k_2)x}, \label{libresol}
\eea 
where $\epsilon_1 = -(k_1+i k_2)^2 , \ k_1> 0, \ k_2 \in {\mathbb R}$. In
general, such a $u_1(x)$ does not tend to zero neither when $x\rightarrow
-\infty$ nor when $x\rightarrow +\infty$. However, two particular
solutions with the required behavior are precisely those of
(\ref{libresol}). We use them for obtaining the nodeless $w(x)$:
\bea
& w(x) = \pm e^{\pm 2k_1 x}/(2k_1).
\eea
It turns out that $\wt V(x)$ becomes again the null potential for both
$w(x)$, $\wt V(x) = 0$. The intermediate 1-SUSY complex potentials
generated by using the two $u_1(x)$ of (\ref{libresol}) are as well
trivial, $V_1(x) = 0$.  Our conclusion is that the null potential can be
non-trivially transformed in frames of our algorithm only at the price of
creating singularities (compare with \cite{BaOvSa95}).

\smallskip

{\it ii)} Consider now the well known one-soliton potential
(P\"oschl-Teller) \cite{DiaNegNieRos99}
\bea
& V(x) = -2 k_0^2 {\rm sech}^2(k_0 x) \label{poschl}
\eea
which is obtained from the null potential by a 1-SUSY transformation
employing $\cosh(k_0 x), \ k_0 >0$. The spectrum of (\ref{poschl}) 
consists of a continuous part $E\geq 0$ and a bound state at $E_0 = -
k_0^2$ with eigenfunction given by: 
\bea
& \psi_0(x) = \sqrt{k_0/2} \ {\rm sech}(k_0 x). \label{gsposchl}
\eea
The general solution $u_1(x)$ of (\ref{schro}) for (\ref{poschl})  with
$\epsilon_1 = -(k_1 + i k_2)^2, \ k_1>0, \ k_2\in{\mathbb R}$ is a linear
combination of the 1-SUSY transformed eigenfunctions of (\ref{libresol})
\bea
& e^{\pm (k_1 + i k_2)x}[k_0 \tanh (k_0 x) \mp (k_1 + i  k_2)].
\label{poschlsol}
\eea
The solutions (\ref{poschlsol}) are precisely the required ones:  if the
upper signs are taken, then $u_1(x) \rightarrow 0$ for $x\rightarrow
-\infty$, while the lower signs ensure $u_1(x) \rightarrow 0$ when
$x\rightarrow +\infty$.  An explicit calculation leads to the two nodeless
$w(x)$: 
\bea
& w(x) = \pm \frac{ke^{\pm 2k_1 x}}{2k_1}\frac{\cosh[k_0(x \mp
x_0)]}{\cosh(k_0 x)}, 
\eea
where $k_0^2 + k_1^2 + k_2^2 \equiv k \cosh(k_0 x_0), \ 2k_0k_1 \equiv
k\sinh(k_0 x_0), \ k,x_0\in{\mathbb R}$. The two 2-SUSY partner potentials
of (\ref{poschl}) read now: 
\bea
& \wt V(x) = -2 k_0^2 {\rm sech}^2[k_0(x \mp x_0)],
\label{npt}
\eea
obtaining just real $x_0$-displaced copies of (\ref{poschl}). This result
has to do with the Darboux invariance phenomenon recently discovered for
the one-soliton well \cite{FerMieRosSam02,FerMieRosSam02b}.

On the other hand, the two complex intermediate potentials generated by
(\ref{poschlsol})  become: 
\bea
& V_1(x) = -2 k_0^2 {\rm sech}^2[k_0(x \mp x_1)], \label{complexposchl}
\eea
where now $k_1 + i k_2 \equiv \kappa \cosh(k_0 x_1), \ k_0 \equiv \kappa
\sinh(k_0 x_1)$ define `complex displacements' $x_1\in{\mathbb C}, \
\kappa\in {\mathbb C}$. These potentials have a bound state at $E_0=
-k_0^2$ whose normalized `ground state' eigenfunction is obtained from
(\ref{eigencomplex}) by employing the $\psi_0(x)$ of (\ref{gsposchl}) and
the $u_1(x)$ of (\ref{poschlsol}):
\bea
\hskip-1cm&& \psi_0^1 (x) \!=\! \frac{k_0}{\vert
\kappa\vert}\bigg[
\frac{1}{k_2}\arctan\big(\frac{k_0 + 
k_1}{k_2}\big) +
\frac{1}{k_2}\arctan\big(\frac{k_0 -
k_1}{k_2}\big) 
\bigg]^{-\frac12} {\rm sech}[k_0(x\mp x_1)]
\eea
The complex potentials (\ref{complexposchl}) were obtained for the first
time in \cite{AndIofCanDed99}.

\smallskip

{\it iii)} Our final example is the harmonic oscillator: 
\bea
& V(x) = x^2, 
\eea
which has a purely discrete equidistant spectrum $\{E_n = 2 n + 1, \
n=0,1,\dots\}$. The general solution of (\ref{schro}) for $\epsilon_1\in
{\mathbb C}$ is now (see \cite{FerHus99} and references therein):
\bea
& u_1(x) = c_1
e^{-\frac{x^2}2}\bigg[{}_1\!F_1\big(\frac{1-\epsilon_1}4,\frac12;x^2\big)
+ 2\nu x\frac{\Gamma(\frac{3-\epsilon_1}4)}{\Gamma(\frac{1-\epsilon_1}4)} 
\, {}_1\!F_1\big(\frac{3-\epsilon_1}4,\frac32;x^2\big)\bigg], \label{hogs}
\eea
where ${}_1F_1(a,c;z)$ is the Kummer hypergeometric series. In general,
such a $u_1(x)$ does not satisfy neither $\lim_{x\rightarrow -\infty}
u_1(x)  = 0$ nor $\lim_{x\rightarrow \infty} u_1(x) = 0$.  However, there
are two particular values for $\nu$ ($\nu = \pm 1$) leading to solutions
with the required behavior:
\bea
& u_1(x) = 
e^{-\frac{x^2}2}\bigg[{}_1\!F_1\big(\frac{1-\epsilon_1}4,\frac12;x^2\big)
\pm 2x\frac{\Gamma(\frac{3-\epsilon_1}4)}{\Gamma(\frac{1-\epsilon_1}4)}
\, {}_1\!F_1\big(\frac{3-\epsilon_1}4,\frac32;x^2\big)\bigg].
\label{goodtfo}
\eea
Take, e.g., (\ref{goodtfo}) with the upper sign, which in the negative
semiaxis $x = -\vert x\vert<0$ reduces to: 
\bea
& u_1(x) = \frac{\Gamma(\frac{3-\epsilon_1}4)}{\Gamma(\frac12)}
e^{-\frac{\vert 
x\vert ^2}2}\Psi\big(\frac{1-\epsilon_1}4,\frac12;\vert x\vert^2\big),
\eea
where the Tricomi function $\Psi(a,c;z)$ is related with ${}_1F_1(a,c;z)$
through (see, e.g., \cite{dekr67}):
\bea
\hskip-1cm & \Psi(a,c;z) = \frac{\Gamma(1-c)}{\Gamma(a-c+1)}{}_1F_1(a,c;z)
+ \frac{\Gamma(c-1)}{\Gamma(a)}z^{1-c}{}_1F_1(a-c+1,2-c;z)
\eea
Since the leading term in the asymptotic expansion for $\Psi(a,c;z)$ is
$z^{-a}$ \cite{AbrSte72}, one can check that $\lim_{x\rightarrow -\infty}
u_1(x)= 0$.  Similarly, if the lower sign is chosen we have
$\lim_{x\rightarrow \infty} u_1(x) = 0$. 

\begin{figure}[ht]
\centering \epsfig{file=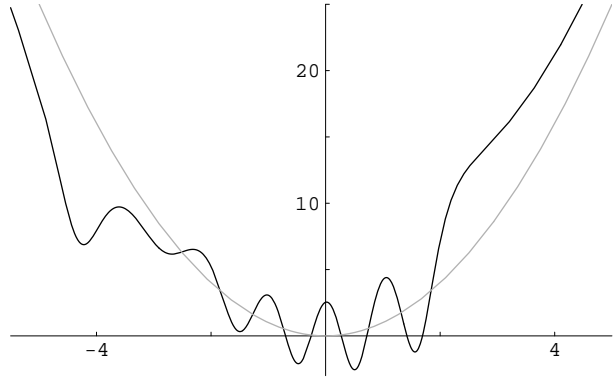, width=8cm}
\caption{\footnotesize The real potential $\wt V(x)$ (black curve)
generated from $V(x) = x^2$ (gray curve) by means of a `complex' 2-SUSY
transformation with $\epsilon_1 = 10 + 0.1 i$ and the $u_1(x)$ of
(\ref{goodtfo}) with the lower minus sign.}
\end{figure}

Once we have identified the solutions (\ref{goodtfo}) with the right
asymptotic behavior, we evaluated $w(x)$ and then $\wt V(x)$. The
resulting expressions in this case are too long; instead, we are plotting
$\wt V(x)$ for $\epsilon_1 = 10 + 0.1 i$ using (\ref{goodtfo}) with the
lower minus sign (see figure 1). Contrasting with the results for the
previous examples, in this case the potentials $\wt V(x)$ are in general
different from $V(x)$. This means that the transformations involving a
pair of complex conjugate factorization energies are effective tools in
generating isospectral 2-SUSY partner potentials. As a byproduct, we have
obtained in a simple way complex potentials $V_1(x)$ given by
(\ref{complexpot}) with real energy eigenvalues $E_n=2n+1$. A plot of the
`ground state' probability density $\vert\psi_0^1(x)\vert^2$, illustrating
the existence of these bound states for the complex 1-SUSY partner
$V_1(x)$ of the oscillator, is shown in figure 2.

\begin{figure}[ht]
\centering \epsfig{file=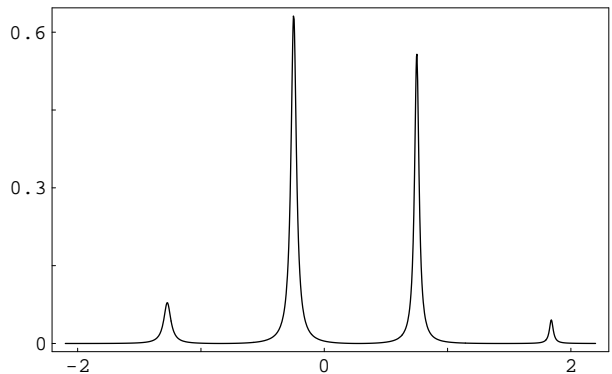, width=8cm}
\caption{\footnotesize The `ground state' probability density
$\vert\psi_0^1(x)\vert^2$ for the complex 1-SUSY partner potential
(\ref{complexpot}) of the oscillator generated by using $u_1(x)$ of
(\ref{goodtfo}) with the lower sign and $\epsilon_1 = 10 + 0.1 i$.}
\end{figure}

\section{Conclusions}

We have shown that the 2-SUSY transformations involving two complex
conjugate factorization energies can produce new non-singular potentials
isospectral to a given initial one. This non-singular character is shared
as well by the intermediate complex potentials arising when those
transformations are factorized. 

{\small 
{\bf Acknowledgements.} 
The support of CONACYT (M\'exico) is acknowledged.  A.R. has been
partially supported by the Spanish Ministerio de Educaci\'on y Cultura
through a FPI grant, research project PB96-0717, and by the European
Commission funding for the Research Training Network \lq\lq Mechanics and
Symmetry in Europe\rq\rq\ (MASIE), contract HPRN-CT-2000-00113.  A.R. 
thanks also the warm hospitality at the Departamento de F\'{\i}sica of
CINVESTAV.}

\end{document}